\documentclass[%
 reprint,
superscriptaddress,
 amsmath,amssymb,
 aps,
 prx,
]{revtex4-2}

\usepackage{amsthm}

\usepackage{graphicx}% Include figure files
\usepackage{dcolumn}% Align table columns on decimal point
\usepackage{bm}% bold math
\usepackage{physics}
\usepackage{booktabs}
\usepackage{enumerate}

\usepackage[pdftex,pagebackref=false]{hyperref} % with 
\hypersetup{
    colorlinks=true,        % false: boxed links; true: colored links
    linkcolor=blue,         % color of internal links
    citecolor=red,        % color of links to bibliography
    filecolor=magenta,      % color of file links
    urlcolor=cyan           % color of external links
}

\usepackage{xcolor}
\usepackage{cleveref}
\usepackage{placeins}
\usepackage{orcidlink}

\newcommand{\iden}{\mathbb{I}}
\newcommand{\diag}{\operatorname{diag}}
\DeclareMathOperator*{\argmax}{argmax}

\newcommand{\ethzurich}{Institute for Theoretical Physics, ETH Z\"urich, 8093 Z\"urich, Switzerland}

\newcommand{\orcidmatteo}{\orcidlink{0000-0002-9426-0377}}
\newcommand{\orcidjannes}{\orcidlink{0000-0001-7491-3660}}
\newcommand{\orcidjuan}{\orcidlink{0000-0001-7263-3462}}

\begin{document}

\title{Majorana string simulation of nonequilibrium dynamics\\ in two-dimensional lattice fermion systems}
\author{Matteo D'Anna$^{\orcidmatteo}$}
\affiliation{\ethzurich}
\author{Jannes Nys$^{\orcidjannes}$}
\affiliation{\ethzurich}
\author{Juan  Carrasquilla$^{\orcidjuan}$}
\affiliation{\ethzurich}

\begin{abstract}
The study of real-time dynamics of fermions remains one of the last frontiers beyond the reach of classical simulations and is key to our understanding of quantum behavior in chemistry and materials, with implications for quantum technology. Here we present a Heisenberg-picture algorithm that propagates observables expressed in a Majorana-string basis using a truncation scheme that preserves Trotter accuracy. 
The framework is exact for quadratic Hamiltonians, where it remains restricted to a fixed low-weight sector determined by the physical observable, it admits variational initial states, and extends to interacting regimes via systematically controlled truncations.
We benchmark our approach on one- and two-dimensional Fermi-Hubbard quenches, comparing against tensor network methods (MPS and fPEPS) and recent experimental data. The method achieves high accuracy on timescales comparable to state-of-the-art variational techniques and experiments, demonstrating that controlled Majorana-string truncation is a practical tool for simulating two-dimensional fermionic dynamics.

\end{abstract}

\maketitle

\section{Introduction}
Real-time dynamics of fermionic lattice models underpin transport, relaxation, and nonequilibrium phase formation in correlated quantum materials.
In two spatial dimensions, direct classical simulation is especially challenging: exact methods scale exponentially and are limited in system size, tensor-network approaches face entanglement growth~\cite{Schollwoeck2011}, dynamical mean-field theory (DMFT)\cite{Georges1996} is approximate away from large coordination, quantum Monte Carlo methods are limited by the sign problem~\cite{Troyer2005}, and neural-network approaches provide a promising yet still developing alternative~\cite{nys2024ab}.
These limitations have motivated both the development of quantum simulators~\cite{fausewehQuantumManybodySimulations2024} and the development of new classical algorithms that push the entanglement frontier~\cite{preskillQuantumComputingEntanglement2012}.   

Ultracold fermions in optical lattices now realize the two-dimensional (2D) Fermi-Hubbard model and provide new insights into phenomena of interacting fermions. Milestones include probing pairing phenomena~\cite{Hirthe2023Nature, Hartke2023Science}, long-range antiferromagnetic order \cite{Mazurenko2017}, time-resolved dynamics of magnetic polarons~\cite{Ji2021PRX}, and the observation of Nagaoka polaron~\cite{Lebrat2024Nature}. Most recently, improvements in cooling and homogeneity have pushed 2D simulators deep into the cryogenic regime at half filling ~\cite{Xu2025Nature}. These advances deliver high-fidelity local observables and quench protocols in regimes that are challenging for numerical simulations, offering a strong incentive for classical methods to provide scalable, quantitatively reliable benchmarks for interpreting analog experiments.

Gate-based devices encode fermions using transformations such as the Jordan-Wigner or Bravyi-Kitaev mappings~\cite{JW1928,BravyiKitaev2002,Seeley2012}, enabling fermionic dynamics to be simulated via product formulas.
Early demonstrations implemented small-scale Hubbard models on superconducting qubits~\cite{Barends2015,Wecker2015}.
Various studies have clarified the role of geometric locality on the simulability~\cite{Campbell2019, cade2020strategies, carobene2024local}, and recent experiments have scaled to 2D Hubbard-like models~\cite{Karamlou2024,Chen2023,Srinivasan2024}.
Current processors remain depth-limited, highlighting the continuing value of classical approaches.

In parallel with hardware progress, classical fermionic simulation continues to advance. Although quadratic Hamiltonians can be simulated efficiently~\cite{TerhalDiVincenzo2002,BravyiFLO2005}, generic interactions break Gaussianity, making numerical simulations challenging due to extensive Heisenberg operator spreading and entanglement growth. Here we present a non-variational, Heisenberg-picture algorithm that operates directly in a Majorana-string basis and pushes the boundaries of simulating geometrically local and parity-preserving lattice Hamiltonians~\cite{bravyi2002fermionic}.
The Majorana propagation (MP) framework we develop here generalizes the recently introduced Pauli propagation (PP, or sparse Pauli dynamics) framework~\cite{Rall2019Simulation, begusic2024Fast, rudolph2025paulipropagationcomputationalframework} to fermionic degrees of freedom. While PP was originally introduced for classical simulations of noisy quantum circuits \cite{Shao2024Simulating,fontanaClassicalSimulationsNoisy2025}, it has since proven broadly applicable. The most relevant applications in the current context range from reproducing utility experiments \cite{rudolph2023classicalsurrogatesimulationquantum} to simulating two- and three-dimensional dynamics \cite{begusic2025realtime}. The connection between PP and MP was also presented in Ref.~\cite{miller2025simulationfermioniccircuitsusing}, which introduced MP to find circuits that approximate ground states of molecular systems.

We develop MP for structured fermionic lattice Hamiltonians and benchmark it on one- and two-dimensional Fermi-Hubbard quenches against matrix product state (MPS) and fermionic projected entangled pair states (fPEPS) calculations, as well as against recent analog Fermi-Hubbard experiments~\cite{Ji2021PRX}. Together, these results show that controlled truncations in the Majorana basis enable accurate estimates of real-time observables in 2D fermionic systems, reaching (and in some regimes surpassing) the accessible time scales of state-of-the-art variational techniques. 

\section{Methods}

\subsection{Majorana basis}
We consider $N$ fermionic modes with fermionic operators $f_i^{(\dagger)}$ that are used to define $2N$ Majorana operators 
\begin{equation}\label{eq:gamma_def}
    \gamma_j=f_j^\dagger+f_j, \quad \gamma_j^{\prime}=i\left(f_j^{\dagger}-f_j\right).
\end{equation}
Majorana operators respect the anticommutation relations
\begin{equation}
    \{ \gamma_i,\gamma_j\}= \{ \gamma_i^{\prime},\gamma_j^{\prime}\} = 2\delta_{ij},\qquad
    \{ \gamma_i^{\prime},\gamma_j\} =0.
\end{equation}
These self-adjoint Majorana operators are used to construct hermitian Majorana strings, represented by the binary vector $v \in \{0,1\}^{2N}$ as \cite{bettaque2025structuremajoranacliffordgroup}
\begin{equation}\label{eq:ms_def}
    \mu(v)\equiv(i)^{v^T \omega_L v} \cdot \gamma_1^{v_1} {\gamma_1^\prime}^{v_2} \cdots \gamma_{N}^{v_{2N-1}}{\gamma_N^\prime}^{v_{2N}}.
\end{equation}
The prefactor $(i)^{v^T \omega_L v}\in\{1,i\}$, with the $2N\times 2N$ lower triangular matrix $(\omega_L)_{ij}\equiv \delta_{i>j}$, recovers the hermiticity of $\mu(v) = \mu(v)^\dagger$. 
Note that the matrix multiplication $v^T \omega_{L} v$ is to be understood as $\operatorname{mod}2$. The Majorana group $\mathcal{M}_{2 N}$ is the set of all strings of length $2N$ 
\begin{equation}
    \mathcal{M}_{2 N}:=\left\{a \cdot \mu(v) \mid a\in\{\pm1,\pm i\}, v \in \{0,1\}^{2N}\right\} .
\end{equation}
Majorana strings provide a basis for (hermitian) fermionic operators directly in the fermionic picture, without the need to specify any fermion-to-qubit mappings, which usually introduce non-local terms.
Since $v$ uniquely determines $\mu(v)$, we will refer to both $v$ and $\mu(v)$ as (Majorana) strings, and start writing fermionic operators using the Majorana basis.
For example, the number operator $n_i$ on site $i$ reads, with the shorthand notation $\mu(v_{ij})\equiv i\gamma_i\gamma_j, \mu(v_{ij^\prime}) \equiv i\gamma_i\gamma_j^\prime$,
\begin{equation}
n_i=f_i^\dagger f_i=\frac{1}{2}\left(1+i \gamma_i \gamma_i^{\prime}\right)=\frac{1}{2}\left(1+\mu\left(v_{ii^\prime}\right)\right).
\end{equation}
Hopping operators between sites $i,j$ (assuming that $i<j$ in the site labeling order chosen in Eq.~\eqref{eq:ms_def}) are written as
\begin{equation}\label{eq:ms_hopping}
f_i^{\dagger} f_j+f_j^{\dagger} f_i=\frac{1}{2}\left(i\gamma_i \gamma_j^{\prime}-i\gamma_i^{\prime} \gamma_j\right)\equiv \frac12\left(\mu(v_{ij^\prime})-\mu(v_{i^\prime j})\right).
\end{equation}
A derivation of these equalities is given in Appendix \ref{app:decompositions}.
$\mathcal{M}_{2 N}$ is closed under multiplication because
\begin{equation}\label{eq:ms_group}
    \mu(v) \mu\left(\tilde v\right)=\zeta\left(v, \tilde v\right) \cdot \mu\left(v+\tilde v\right),
\end{equation}
where $v+\tilde v$ is to be understood as the bitwise addition $\operatorname{mod}2$ and, for $\omega \equiv \omega_L + \omega_L^T$,
\begin{equation}
    \zeta\left(v, \tilde v\right) = (-1)^{g(v,\tilde v)}i^{v^T \omega \tilde v} \in \{\pm1,\pm i\},
\end{equation}
where
\begin{equation}
\begin{split}
    g\left(v, \tilde v\right) \equiv~&v^T \omega_L \tilde v+ (v^T \omega_L v)(\tilde v^T \omega_L \tilde v) \\
    + &v^T \omega \tilde v(v^T \omega_L v+\tilde v^T \omega_L \tilde v+1).
\end{split}
\end{equation}
More importantly for MP, the strings satisfy the (anti)commutation relation
\begin{equation}\label{eq:ms_commrel}
    \mu(v) \mu\left(\tilde v\right)=(-1)^{v^T \omega \tilde v} \mu\left(\tilde v\right) \mu(v).
\end{equation}
Commutation or anticommutation between two strings $v, \tilde v$ is therefore fully determined by the value of $v^T \omega \tilde v\in\{0,1\}$.

For any Majorana string $\mu(v)$ we define the weight $w(v) \in\{0,\cdots, 2N\}$ as the number of unique Majorana operators in $\mu(v)$ or, equivalently, as the number of non-zero entries in $v$, i.e.\
\begin{equation}\label{eq:ms_weight}
w(v) = \sum_{i=1}^{2N}v_i.
\end{equation}
The fermion parity $p(v)$ is defined as $p(v)=(-1)^{w(v)}$.
Hermitian Majorana strings describing physical observables must commute with the parity operator in order to respect fermionic superselection rules, and this is only satisfied by even parity strings, i.e., strings with an even number of Majorana operators.
A Majorana mode $i$ is said to be \emph{unpaired} if a string contains Majorana operator $\gamma_i$ without $\gamma_i'$, or vice versa, i.e.\ $v_{2i-1} \neq v_{2i}$. The number of unpaired Majorana operators in the string $\mu(v)$ is denoted by $w_s(v) \in \{0,\cdots, N\}$
\begin{equation}
    w_s(v) = N-\left(\sum_{i=1}^N \delta_{v_{2i-1}, v_{2i}}\right). \label{eq:singles_count}
\end{equation}

In Appendix \ref{app:decompositions} we show the simple but informative relation for the phase factor in Eq.~\eqref{eq:ms_commrel}
\begin{equation}\label{eq:bitstrings_comm_main}
    v^T\omega \tilde v = \left(w(v)w(\tilde v) - w(v \odot \tilde v)\right)\operatorname{mod}2,
\end{equation}
where $\odot$ is elementwise multiplication. With Eq.~\eqref{eq:bitstrings_comm_main}, we find that such even parity strings commute if they have an even number of Majorana operators in common, and they anticommute if they have an odd number in common. This relates to the fact that $w(v)w(\tilde v)$ is always even for even parity strings, and hence $w(v \odot \tilde v)$ determines the parity in Eq.~\eqref{eq:bitstrings_comm_main}.

\subsection{Majorana propagation}
We focus on simulating the dynamics of fermionic lattice Hamiltonians.
Given an initial state $\rho$ and an observable $O$, we aim to evaluate the expectation value
$\Tr \left( U(\tau) \rho U^{\dagger}(\tau) O \right)$, where $U(\tau)$ denotes the time-evolution operator over a duration $\tau$.
As within PP, we switch to the Heisenberg picture, and write $O$ as a linear combination of Majorana strings $O = \sum_v \alpha_v \mu(v)$, such that
\begin{equation}\label{eq:mp_ev}
    \Tr\left(U(\tau)\rho U(\tau)^{\dagger} O\right) = \sum_v \lambda_v \Tr\left(\rho \mu(v)\right).
\end{equation}
with $\lambda_v\in\mathbb R$ a real-valued coefficient to each string that we aim to determine using MP. We write the unitary dynamics as a sequence of gates (i.e.\ using Trotterization) $U(\tau)= U_m\cdots U_1$, where each gate $U_m$ corresponds to the rotation with generator string $v^m \in\{0,1\}^{2N}$, $U_m=\mathrm{exp}(-i\theta_m \mu(v^m)/2)$. We obtain the coefficients $\lambda_v$ using
\begin{equation}
    \begin{split}
    U(\tau)^{\dagger} & O U(\tau)  = U_1^\dagger \cdots U_m^\dagger\left(\sum_v \alpha_v\mu(v)\right)U_m\cdots U_1 \\
    & = U_1^\dagger \cdots U_{m-1}^\dagger \left(\sum_v \lambda^{(m)}_v\mu(v)\right)U_{m-1}\cdots U_1 \label{eq:linearcomb} \\
    &~\vdots \\
    & = \sum_v \lambda_v\mu(v) 
\end{split}
\end{equation}
To compute the coefficients $\lambda^{(m)}_v$ after applying the $m$'th unitary $U_m$ we write it as
\[\exp(-i\theta\mu(v^m)/2) = \cos(\theta/2) \iden -i\sin(\theta/2) \mu(v^m),\]
and find the following \emph{MP splitting rule} (analogous to splitting rule in PP)
\begin{equation}
\label{eq:mp_splitting}
\begin{aligned}
&e^{i \tfrac{\theta}{2} \mu(v^m)}
\,\mu(v)\,
e^{-i \tfrac{\theta}{2} \mu(v^m)}=
\\
&=
\begin{cases}
\mu(v)
& \text{if } [\mu(v^m), \mu(v)] = 0,
\\[4pt]
\begin{aligned}
&\quad\cos(\theta)\mu(v) \\
&+s\cdot \sin(\theta)
\,\mu\bigl(v + v^m\bigr)
\end{aligned}
& \text{if } \{\mu(v^m), \mu(v)\} = 0.
\end{cases}
\end{aligned}
\end{equation}
where $v+v^m$ again corresponds to a new string obtained through bitwise $\operatorname{mod}2$ addition, and the sign $s=\pm1$ in the anticommuting case is given by 
\begin{equation} 
s\equiv (-1)^{g(v, v^m)}.
\end{equation}
We hence find that applying Majorana rotations to Majorana strings in MP follows the same scheme as in PP, with the adjustment that the (anti)commutation of $v,v^m$ is determined by Eq.~\eqref{eq:ms_commrel} and not by the commutation relations of Pauli strings as in PP. We loosely refer to the case of anticommutation as the ``splitting branch'', since it causes an increase in the total number of Majorana strings in the observable, $v \to v, v+v^m$. 

The splitting branch is the reason why Majorana propagation for a general Hamiltonian will inevitably face a computational barrier due to the increasing number of strings with non-negligible $\lambda_v$. However, below we discuss two special cases where the simulation avoids this computational barrier.

First, unitaries consisting solely of fermionic Clifford gates, corresponding to the rotation $\exp(-i\theta\mu(v^m)/2)$ with rotation angles $\theta\in\{k \pi/2, k\in\mathbb N\}$ \cite{bettaque2025structuremajoranacliffordgroup}, can be simulated efficiently with Majorana propagation. This can be easily understood from the splitting rule in Eq.~\eqref{eq:mp_splitting}, where the ``splitting branch'' does not generate additional strings.

\begin{figure}
    \centering
    \includegraphics[width=0.99\linewidth]{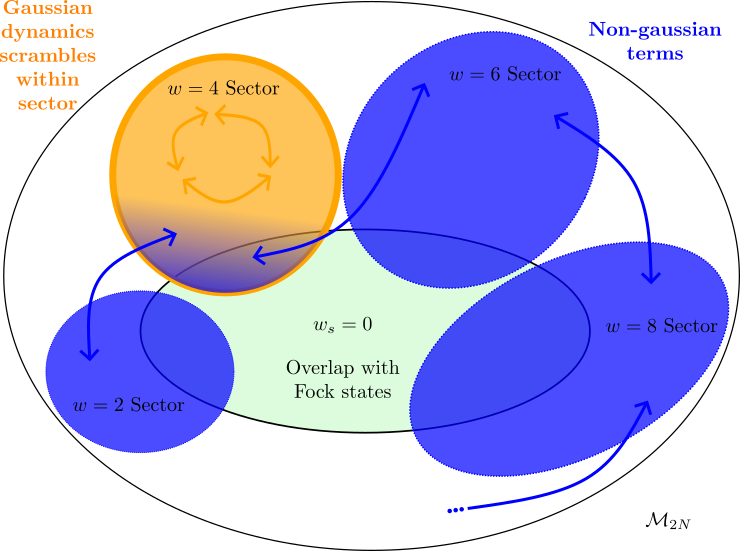}
    \caption{Gaussian dynamics (orange sector) is restricted to the weight sector(s) of the Majorana strings that are chosen to be evolved. In contrast, evolution under non-Gaussian dynamics (red sectors) generates strings in all weight sectors.}
    \label{fig:sectors}
\end{figure}

A second, more important exception is the simulation of fermionic Gaussian dynamics, or the dynamics governed by a quadratic Hamiltonian
\begin{align}
H= \sum_{ij} f_i^\dagger h_{ij} f_j .
\end{align}
This is less obvious, and is proven in Appendix~\ref{sec:gauss_dynamics}, where we show that the dynamics is constrained to a low-weight subspace of Majorana strings determined by the observable $O$. Hence, as long as the subspace of a given weight $w$ is small enough (as for the spatially local observables considered here) the dynamics is classically efficient. For more details see Appendix~\ref{sec:gauss_dynamics}.
This ``weight conservation'' can be understood as follows. New strings are solely generated through the sine branch in \eqref{eq:mp_splitting}. For a unitary generated by a weight $2$ Majorana string $\mu(v_{ij^\prime})$ (e.g.\ a hopping unitary) applied to a string $\mu(v)$, they are of the form $v\to v'=v+v_{ij^\prime}$. However, at the same time, to activate the sine branch, $\mu(v)$ must anticommute $\mu(v_{ij^\prime})$, which by Eq.~\eqref{eq:bitstrings_comm_main} means that they share exactly one non-zero index: $w(v \odot v_{ij^\prime}) = 1$. Since we have $w(v_{ij^\prime})=2$ and in general $w(v+\tilde v) = w(v) + w(\tilde v) - 2w(v \odot \tilde v)$, we obtain $w(v) = w(v+v_{ij^\prime})$, showing that the weight is indeed unaltered, see Figure \ref{fig:sectors}.
Hence, from this observation, we can regard non-Gaussianity of the Hamiltonian (rather than the physically less relevant Clifford structure) as a degree of complexity for Majorana propagation.

\subsection{Truncations}
To mitigate exponential growth in the number of strings of the observable while retaining accuracy, we employ truncation schemes to reduce the number of Majorana strings. Namely, we perform two truncations to mitigate the growth of Majorana strings:
\begin{enumerate}
    \item The first truncation is \textit{coefficient truncation}: whenever a string $\mu(v)$ in the linear combination in Eq.~\eqref{eq:linearcomb} has $|\lambda_v^m| < \varepsilon$, for a fixed $\varepsilon$, we remove $\mu(v)$ from the linear combination.

    \item The second truncation is based on the number of unpaired Majoranas $w_s(v)$ in Eq.~\eqref{eq:singles_count}. We refer to it as \textit{unpaired truncation}.
    This truncation, which is especially well motivated if the initial state is a Fock state, consists of truncating strings $\mu(v)$ with $w_s(v) > S\in\mathbb N^+$. We provide a detailed motivation for this truncation below.
\end{enumerate}
We will detail the reasoning behind these truncation rules in the next subsections.

\subsubsection{Motivation of the unpaired truncation}
To motivate our choice of truncating string based on $w_s(v)$, we observe that only Majorana strings $\mu(v)$ with fully paired Majoranas have a non-vanishing overlap with Fock states
\begin{equation}\label{eq:fock_state}
\left|n_1 \cdots n_N\right\rangle:=\big(f_N^{\dagger}\big)^{n_N} \cdots\big(f_1^{\dagger}\big)^{n_1}|0\rangle,
\end{equation}
for $|0\rangle$ the fermionic vacuum.
In fact, it holds
\begin{equation}\label{eq:Fock_overlap}
\begin{split}
    \left\langle n_1 \cdots n_N\right | & \mu(v)\left|n_1 \cdots n_N\right\rangle  = \\ &= (i)^{v^T \omega_L v} \prod_{j=1}^N  \left(i (-1)^{n_j}\right)^{v_{2j}} \delta_{v_{2j-1},v_{2j}}.
    \end{split}
\end{equation}
A detailed derivation of this expression is given in Appendix \ref{sec:Fock_overlap}.
Notice that the amplitude of the overlap is the same for all strings $v$ with $w_s(v)=0$: $\abs{\left\langle n_1 \cdots n_N\right | \mu(v)\left|n_1 \cdots n_N\right\rangle} = 1$.

Therefore, because of \eqref{eq:Fock_overlap}, as we propagate further towards the initial state, it becomes increasingly unlikely that strings $v$ with a large $w_s(v)$ will contribute significantly to the final expectation value. 
A simple truncation rule based on estimations of the potential overlap with the initial state is not always available for a general initial state $\rho$ beyond a pure Fock state. For simplicity, we hence restrict ourselves to initial states close to pure Fock states. 

\subsubsection{Connection of unpaired truncation with Trotterization}
To maintain high accuracy, it is important that the errors introduced by truncating Majorana strings are consistent with those introduced by the Trotterization scheme. Although this has not been pointed out in the context of Pauli propagation, we introduce the concept of Trotter-consistent cutting, which is also directly applicable to the former framework.

Consider a single $p$-th order Trotter layer $U(\tau)$ such that
\begin{align}\label{eq:trotter_order}
    U(\delta\tau) = e^{-i H \delta \tau} + \order{(\delta \tau)^{p+1}}.
\end{align}
To analyze the Heisenberg action on an observable $O$, we apply Campbell's identity \cite{CampId}
\begin{equation}\label{eq:Camp_id_clean}
    U^\dagger(\delta\tau) O U(\delta\tau)
    =
    \sum_{k=0}^{p} \frac{(i \delta\tau)^k}{k!} [H,O]_k
    + \order{(\delta\tau)^{p+1}},
\end{equation}
where we defined the nested commutators
\begin{equation}
    [X, Y]_k = \begin{cases}
        Y & \text{if } k=0 \\ \underbrace{[X, \cdots[X,[X}_{k \text { times }}, Y]] \cdots] & \text{if } k> 0,
    \end{cases}
\end{equation}
We now split the Hamiltonian into parts that preserve and alter the number of unpaired Majorana operators
\begin{equation}
    H = H_P + H_{NP}.
\end{equation}
Here, $H_P$ preserves the number of unpaired Majoranas (e.g., density or interaction terms) and $H_{NP}$ is the non-preserving part (e.g., hopping terms). At order $k$ in \eqref{eq:Camp_id_clean}, the largest change in the number of unpaired Majorana operators arises from the contribution with $k$ commutators by $H_{NP}$, i.e.\ $[H_{NP}, O]_k$.
If $H_{NP}$ is composed of hopping terms, each commutator can change the number of unpaired Majoranas by at most $2$. Consequently, 
\begin{align}
\Delta w_s(v) \in \{+2, 0, -2\}.
\end{align}
This observation motivates a truncation rule that is consistent with the Trotter error budget. 

\subsubsection{Trotter-consistent truncation scheme}\label{sec:wstrunc}
Since the Trotter formula already neglects terms of order $(\delta\tau)^{p+1}$, it is unnecessary to retain MP contributions whose effect would only appear at higher orders than $(\delta\tau)^{p}$. Accordingly, we impose the following truncation strategy.
\begin{itemize}
\item Fix a global cap $S$ on the number of unpaired Majoranas retained after each Trotter layer. After completing a Trotter layer, discard all strings with $w_s(v) > S$.
\item Within a Trotter layer, allow \textit{temporary} growth up to $S' = S + (2p)/2 = S + p$, corresponding to the maximal allowed order in $\delta\tau$ generated by $p$ nested commutators with $H_{NP}$.
\end{itemize}
This Trotter consistent cutting ensures that strings which can recombine through the nested commutator structure of the $p$-th order expansion into terms with $w_s \le S$ are not prematurely discarded, while contributions whose effects would only appear beyond order $(\delta\tau)^p$ are consistently neglected together with the intrinsic Trotter error. Coefficient truncation is enforced after each gate application. 

\subsubsection{Relation to other truncation schemes}
A typical truncation in PP is \textit{weight truncation} (also used for MP in Ref.~\cite{miller2025simulationfermioniccircuitsusing}). Instead of counting the number of unpaired operators $w_s(v)$, weight truncation is imposed on the total weight $w(v)$ in Eq.~\eqref{eq:ms_weight}, i.e.\ one truncates strings $v$ with $w(v) > W$.
However, $w_s(v)$ is a better indicator for estimating the overlap of a Majorana string with an initial Fock state, as discussed above.
In the context of PP, the equivalent of the unpaired truncation is the ``$X$-truncation'' introduced by \cite{begusic2025realtime}.

\subsubsection{Coefficient truncation}
An upper bound on the number of generated strings can be obtained by analyzing the norm of the observable. Assume we write the observable after $K$ steps as $O_K = \sum_v \lambda_v \mu(v)$. 

Unitary evolution strictly conserves the norm $||O_K||_2^2 = ||O_0||_2^2$, while coefficient truncation decreases it. Since every surviving string contributes at least $\varepsilon^2$, the number of surviving strings $M$ is bounded by
\begin{equation}\label{eq:2norm}
M \leq ||O_0||_2^2 / \varepsilon^2.
\end{equation}

On the other hand, the 1-norm $||O||_1 = \sum_v |\lambda_v|$ allows to capture the dependence on the circuit depth $K$. At each gate application, a string either commutes (retaining its coefficient) or anticommutes (splitting into the cosine and the sine branch). In the latter case, the 1-norm of the coefficients involved in the splitting grows by a factor $\xi = |\cos(\theta)| + |\sin(\theta)|$. Since $\xi \geq 1$, the worst-case total 1-norm growth assumes every string splits at every step, and in this case, we have 
\begin{equation}
||O_{K}||_1 \leq \xi ||O_0||_1.
\end{equation}
Let $p_1,\cdots, p_l$ be $l$ different physical parameters needed to fully describe the system. Then the rotation angles of the various gates will be of the form $\theta_j = p_j \delta\tau$. Thus the tracked observable after $K=K_1+\cdots + K_l$ steps, for $K_j$ to the number of times a gate with parameter $\theta_j$ was applied, satisfies 
\begin{equation}
||O_K||_1 \leq \left(\prod_{j=1}^l \xi_j^{K_j}\right) ||O_0||_1 \equiv \Xi ||O_0||_1.
\end{equation}
for $\xi_j = |\cos(\theta_j)| + |\sin(\theta_j)|$.
Since every surviving string must contribute at least $\varepsilon$ to the 1-norm, we find
\begin{equation}\label{eq:1norm}
    M \leq \Xi||O_0||_1 / \varepsilon.
\end{equation}

The number of strings is hence upper-bounded by the minimum of Equations \eqref{eq:2norm} and \eqref{eq:1norm}.
Since for small time steps, $|\cos(\theta_j)| + |\sin(\theta_j)| = 1 + |p_j| \delta \tau + \order{(|p_j| \delta \tau)^2}$, we have
$\left(|\cos(\theta_j)| + |\sin(\theta_j)|\right)^{K_j} \approx e^{K_j |p_j| \delta\tau}$. We hence find
\begin{equation}\label{eq:Mbound}
    M \leq \min\left( \frac{||O_0||_2^2}{\varepsilon^2}, \frac{||O_0||_1 e^{\delta \tau \left(\sum_{j=1}^l K_j |p_j|\right)}}{\varepsilon} \right).
\end{equation}

\section{Numerical results}

\subsection{Gaussian dynamics}
\begin{figure}
    \includegraphics[width=.49\textwidth]{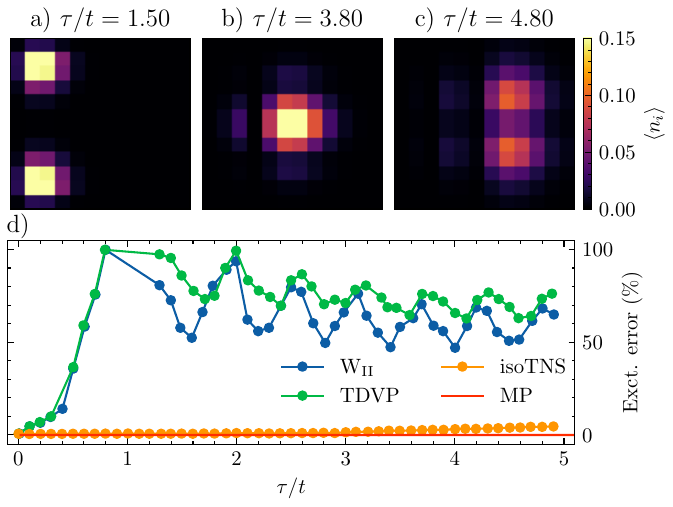}
    \caption{Interference dynamics of two spinless free fermions on a $12\times 12$ lattice at $U/t=0$. a-c) Plot of the local densities before ($\tau/t=1.50$), during ($\tau/t=3.80$) and after ($\tau/t=4.80$) the scattering process, using a time-step $\delta \tau/t = 0.01$. MP reproduces the dynamics exactly. d) Excitation error $\sum_{i \in \mathcal I_e} \Delta n_i / n_{\text {tot }}$ introduced in Ref.~\cite{dai2025fermionic} with $\eta=0.6$ (see \eqref{eq:exc_sites} for the definition of the excited sites $\mathcal I_e$ and $\eta$), compared with the data reported in Ref.~\cite{dai2025fermionic} for $\mathrm{W}_{\mathrm{II}}$~\cite{zaletel2015time} and with the time-dependnt variational principle (TDVP)~\cite{haegeman2016unifying} using MPS, and 2D isoTNS. The excitation error for Majorana propagation is mainly determined by the Trotter error, and can be made arbitrarily small.}\label{fig:scattU0}
\end{figure}

As a first test of the capabilities of our approach, we focus on one model employed in Ref.~\cite{dai2025fermionic}: the (spinless) free-fermion model
\begin{equation}\label{eq:spinless}
     H_\mathrm{h}=-t\sum_{\langle i, j\rangle} f_i^{\dagger} f_j+f_j^{\dagger} f_i.
\end{equation}
As in Ref.~\cite{dai2025fermionic}, we study the interference patterns of two fermions initially placed on adjacent corners of a 2D square lattice with open boundary conditions that are left to evolve under Eq.~\eqref{eq:spinless}. The local densities at different stages of the process on a $12\times 12$ lattice are reported in Figure~\ref{fig:scattU0}.

Free-fermion scattering in two dimensions can be simulated exactly and efficiently using Gaussian-dynamics techniques~\cite{surace2022fermionicgaussian}. In contrast, this task remains challenging for several widely used classical variational approaches, such as matrix product state (MPS) methods (see Ref.~\cite{dai2025fermionic} and benchmarks therein~\cite{haegeman2016unifying, zaletel2015time}). In our comparisons in Figure \ref{fig:scattU0}, we use standard MPS techniques, noting that their performance could be further improved by incorporating mode-transformation schemes~\cite{krumnow2016fermionic, krumnow2019towards}. Recently proposed fermionic PEPS (fPEPS) methods~\cite{dai2025fermionic} achieve much smaller errors, though these still increase gradually over time.

In contrast, within the Majorana-propagation framework, the noninteracting regime can be simulated efficiently, without any truncation in $w_s$. As detailed in Appendix~\ref{sec:gauss_dynamics}, the computation remains confined to a subspace of strings with weight $w$ determined by the structure of the observable $O$. In particular, the dynamics of local densities is fully captured within the $w(v)=2$ subspace of Majorana strings.

\subsection{Interacting Fermi-Hubbard dynamics: 1D}\label{sec:1dFH}
Moving on to challenging dynamics with interacting fermions, we consider the Fermi-Hubbard model
\begin{align}
    H &=-t\sum_{\langle i,j\rangle, \sigma \in\{\uparrow, \downarrow\}}\left(f_{i, \sigma}^{\dagger} f_{j, \sigma}+f_{j, \sigma}^{\dagger} f_{i, \sigma}\right)+U\sum_i n_{i \uparrow} n_{i \downarrow} \label{eq:fh_ham} \\
    &= H_\mathrm{h} + H_\mathrm{r}  \nonumber 
\end{align}
We first consider a $1$D chain of $100$ spinful sites, where we simulate the dynamics of an initial ferromagnetic eigenstate at $U/t=+\infty$ evolving under the quenched Hamiltonian with $U/t=1$.  We implement a second-order Trotter expansion
\begin{equation}\label{eq:2ndorderTrotter}
    U(\delta \tau) = e^{-i\frac12\delta \tau H_\mathrm{h}} e^{-i\delta \tau H_\mathrm{r}} e^{-i\frac12\delta \tau H_\mathrm{h}}.
\end{equation}
where gates in the third factor are applied in reverse order.
The expressions of the hopping and repulsion gates in terms of Majorana operators are listed in Appendix~\ref{app:decompositions}. Since we use a second-order expansion, our Trotter-consistent unpaired truncation technique imposes a relaxed cutoff $S'=S+2$ within each Trotter layer in all experiments.
As already remarked above, the hopping terms in the Hamiltonian change the number of unpaired operators, while leaving the weight of the string unchanged. On the other hand, the repulsion term $\mu(v_{ii^\prime jj^\prime})\equiv \gamma_i\gamma_i^\prime\gamma_j\gamma_j^\prime$ increases the weight of the strings, but preserves the number of unpaired operators.

\begin{figure}
    \includegraphics[width=.48\textwidth]{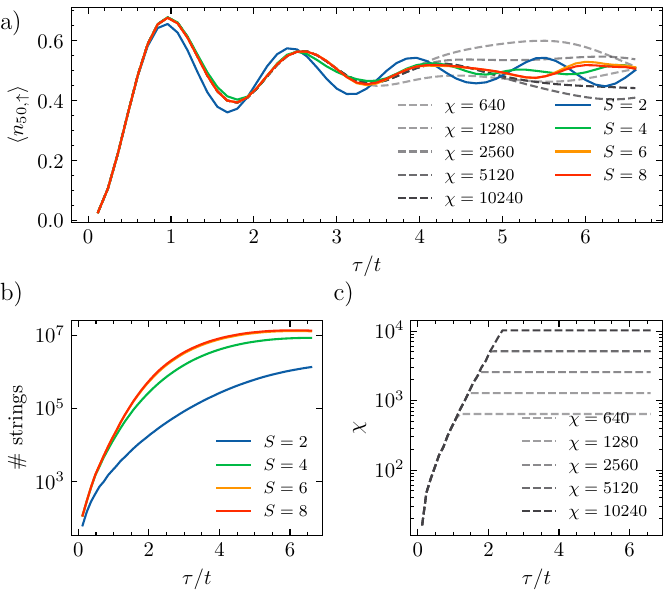}
    \caption{a) Expectation value of the density $n_{50,\uparrow}$ at a central site for a $1$D lattice with $100$ (spinful) sites, subject to second-order Trotter dynamics with $\delta\tau=0.12$. We compare the results from Majorana propagation for the unpaired truncation at various cutoffs $S$ (solid colored lines) with MPS of different bond dimensions (dashed grey lines). The coefficient truncation for all Majorana propagation simulations is $\varepsilon=10^{-5}$. b) Growth of the number of Majorana strings with time at various cutoffs $S$. c) Maximum bond dimension obtained with the MPS simulations.}
    \label{fig:QA_1d}
\end{figure}

In Figure \ref{fig:QA_1d} we show the local densities at the central site $n_{50,\uparrow}(\tau)$ at different cutoffs $S$ (and $\varepsilon=10^{-5}$). We compare the results with those of a fermionic MPS with various bond dimensions $\chi$ that evolves under the same Trotter circuit.
Even with the strong coefficient truncation, the Majorana propagation is accurate to $\tau/t\approx 4$. Furthermore, we observe little qualitative difference in the predictions between $S=6$ and $S=8$ over the entire time interval, and the collapse further indicates that $S \geq 6$ yields accurate predictions. In contrast, the MPS predictions for the same Trotter circuit diverge as $\tau/t \approx 5$ for the considered bond dimensions. 

\begin{figure*}
    \includegraphics[width=\textwidth]{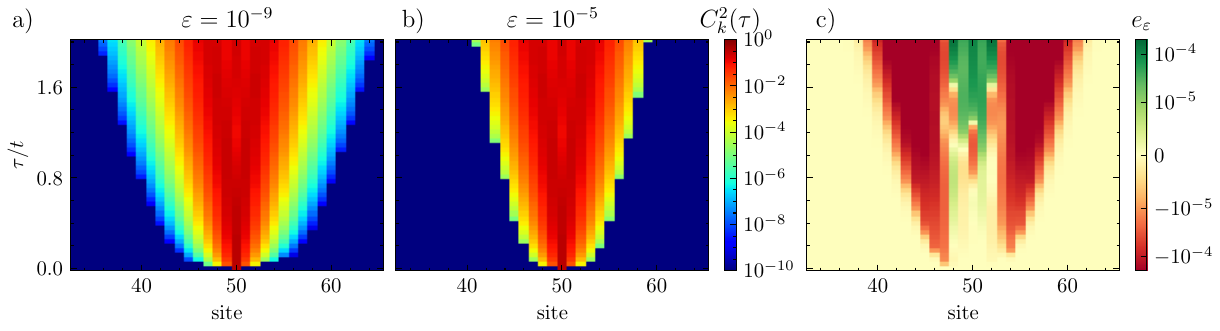}
    \caption{Space-time map of the OTOC $C^2_k(\tau):= ||[n_{50,\uparrow}(\tau), L_k]||$ of $n_{50,\uparrow}(\tau)$ with the local operator $L_k$ on a 100 site 1D chain. Panels a) and b) show the results for different coefficient truncations $\varepsilon = 10^{-9}$ (a) and $\varepsilon = 10^{-5}$ (b). In both simulations $S=4$. c) Plot of the OTOC differences $e_\varepsilon$ for $\varepsilon=10^{-5}$ against the most accurate result $\varepsilon=10^{-9}$. The full 1D lattice consists of 100 sites. }\label{fig:1dLR}
\end{figure*}

More generally, we find that at short to intermediate time scales, accurate results are obtained by setting a small fixed $\varepsilon \approx 10^{-5}$, and optimizing $S$ until the observable predictions converge. The growing number of Majorana strings is shown in Fig.~\ref{fig:QA_1d} (b), where we can observe the effect of the cutoff $S$. At larger time scales, smaller $\varepsilon$ can become relevant, but introduce a prohibitive computational cost. 
The coefficient truncation is a tool for taming the growth in the number of strings, also shown in Fig.~\ref{fig:QA_1d} (b). Here, we numerically study the impact of this truncation on the dynamics by measuring local infinite-temperature out-of-time-order correlators (OTOC) and visualizing the Lieb-Robinson (LR) bound. After each Trotter layer, we compute the commutator of the time-evolved observable $O(\tau)$ at time $\tau$ with an (arbitrary) local operator at site $k$
 \begin{equation}\label{eq:Lk}
 L_k = \sum_{\sigma\in \{\uparrow, \downarrow\}} i\gamma_{k,\sigma}\gamma^\prime_{k,\sigma} + \gamma_{k,\sigma}+\gamma^\prime_{k,\sigma}.
 \end{equation}
The LR bound imposes a finite information propagation speed, such that Eq.~\eqref{eq:Lk} decays exponentially with distance outside the light cone, i.e.\ for some $c, K,\tilde K, \nu\in \mathbb R$ \cite{lieb1972finite, chen2023speedlimits}
\begin{equation}
\begin{split}
    &\left(\Tr\left([O(\tau), L_k]^\dagger [O(\tau), L_k]\right)\right)^{\frac12} \\
    &=\norm{[O(\tau), L_k]} \\
    &\leq K \frac{(c\tau)^r}{r!} \leq \tilde K e^{\nu(vt -r)}
\end{split}
\end{equation}
where $r$ is the distance (in terms of lattice units) between the initially localized observable $O(0)$ and site $k$ and $v=e^\nu c/\nu$. In Figure \ref{fig:1dLR} we show $\norm{[O'(\tau), L_k]}$ for different choices of $\varepsilon$, and observe that the coefficient truncation shortens the tails of the information profile. It affects the OTOC norm within the light cone and the interference patterns, especially at shorter distances.

\begin{figure*}
    \includegraphics[width=\textwidth]{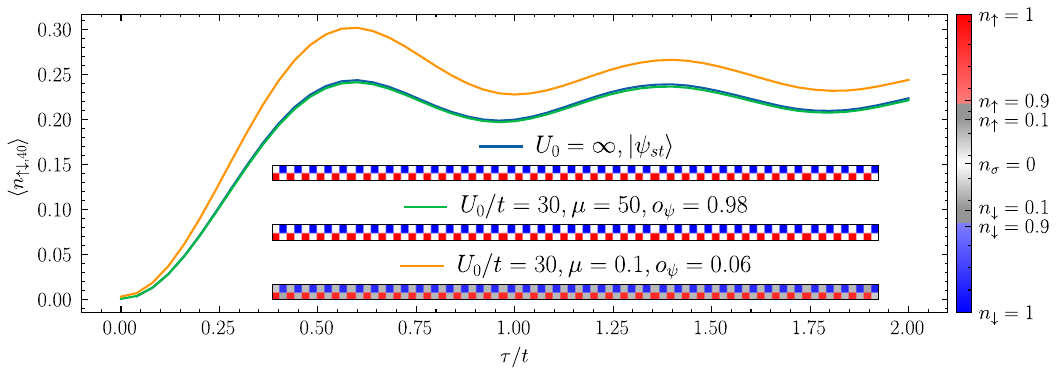}
    \caption{Expectation values of the doublon occupation probability $n_{40,\uparrow\downarrow}$ for a ground state $\ket\psi$ prepared with $U_0/t=30$ and the different shifts in chemical potential $\mu=50, 0.1$, that is then quenched with the Fermi-Hubbard Hamiltonian with $U/t=1$. The staggered state $\ket{\psi_{st}}:=\ket{\uparrow\downarrow\cdots\uparrow\downarrow}$ corresponds to the ground state of $U_0/t=\infty, \mu> 0$. For the overlap of ground states $\ket\psi$ at finite $U_0,\mu$ with $\ket{\psi_{st}}$ we use the notation $o_\psi = |\braket{\psi}{\psi_{st}}|^2$. The MP truncations are $S=10, \varepsilon=5\cdot 10^{-6}$.}\label{fig:1dmps}
\end{figure*}

We now extend the approach to quenches where the initial state is a variational approximation of the ground state for a finite $U_0/t > 0$. Notice that without interaction, the initial state could be absorbed into Majorana propagation via a fermionic Gaussian unitary on a single Fock basis state, as shown in Appendix~\ref{app:gaussian_initial}. Here, however, we use DMRG~\cite{dmrg2005} to represent the interacting ground state of some Hamiltonian of interest (specified below) as a MPS $\ket\psi$, and evaluate $\langle \psi | \mu(v)|\psi\rangle$ in Eq.~\eqref{eq:mp_ev} by translating $\mu(v)$ into an MPO. 
We report the results for a 1D chain of $80$ sites under the Fermi-Hubbard Hamiltonian in Eq.~\eqref{eq:fh_ham}. The initial state is prepared as the ground state at a strong interaction $U_0/t=30$ while adding an alternating local chemical potential 
\begin{align}
H_{\mu} &= - \mu \sum_{i} n_{i, \uparrow} \delta_{i, \text{even}} + n_{i, \downarrow} \delta_{i, \text{odd}}
\end{align}
at half filling.
In Figure~\ref{fig:1dmps} we plot the doublon density $n_{i,\uparrow\downarrow}$ at the site $i=40$ for two situations: for $\mu=50$ and $\mu=0.1$, such that $o_\psi = |\braket{\psi}{\psi_{st}}|^2 = 0.98$ and $0.06$ respectively, where $\ket{\psi_{st}}\equiv\ket{\uparrow\downarrow\cdots\uparrow\downarrow}$ is the staggered ground state at 
$U_0/t=\infty,\mu>0$ (also reported in Figure \ref{fig:1dmps} for comparison).
For $\mu=0.1$, the dynamics behave quantitatively differently, yielding increased hole probabilities throughout the evolution. Establishing useful use cases and rigorous bounds for such ``mixed picture'' approaches is an important avenue for future development of Heisenberg picture methods, where MP (and PP) can complement other classical methods, and vice versa. In Figure \ref{fig:mpshist} in Appendix \ref{app:additional_experiments} we report the overlaps $\Tr(\rho \mu(v))$ with the initial states. Notice that strings with unpaired operators now have non-zero overlap with the initial state, in contrast to Eq.~\eqref{eq:Fock_overlap}. For $\mu=0.1$ especially, there is a significantly increased contribution from higher $w_s(v)$. Developing dedicated truncation rules for generic ground states is an example of such avenues of research.

\subsection{Interacting Fermi-Hubbard dynamics: 2D}
We investigate the applicability of the MP method in simulating modern quantum analog experiments at strong interactions. We study the real-time dynamics of a two-dimensional Fermi-Hubbard system initialized in an antiferromagnetic state doped with a single hole at the lattice center.  This setting captures the subtle interplay between spin and charge degrees of freedom, whose coupling gives rise to magnetic polarons, which are mobile charge excitations dressed by local spin distortions, believed to underlie emergent phenomena such as high-temperature superconductivity. The same scenario has been realized in state-of-the-art ultracold-atom quantum simulators \cite{Ji2021PRX}, making it an ideal benchmark for validating our MP approach against real experimental observations of polaron formation.

\begin{figure*}
    \includegraphics[width=\textwidth]{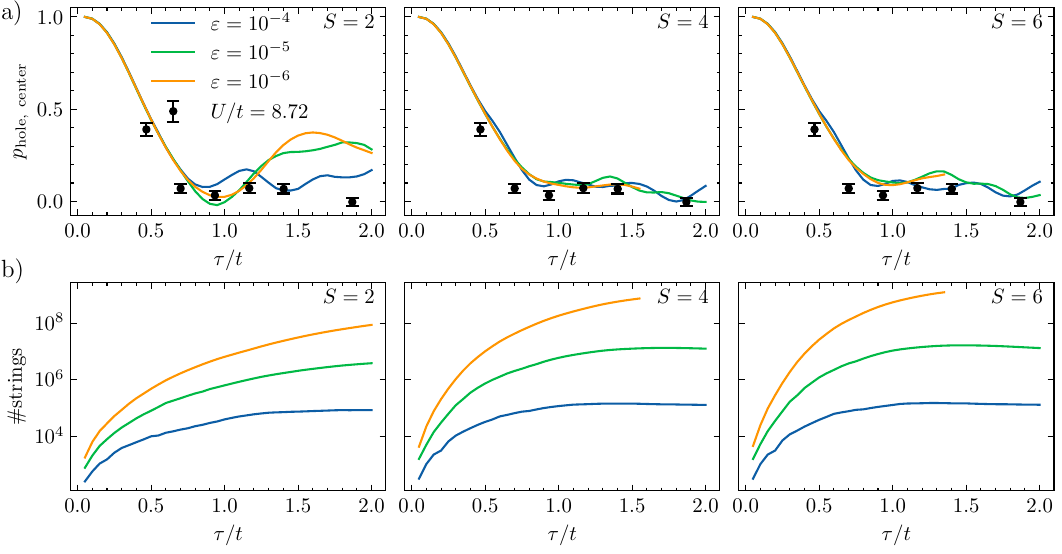}
    \caption{a) Comparison of $p_\mathrm{hole}(\tau)$ obtained with Majorana propagation on a $7\times 7$ lattice compared to the experimental results reported in Ref.~\cite{Ji2021PRX}. As in the experiment, we set $U/t=8.72$. The three columns report the results for the truncations $S=2,4,6$, respectively. For each choice of $S$, we truncate at different values of $\varepsilon$. For $S=4,6,$ we observe good qualitative agreement between the experimental and simulated data. b) The number of strings in the observables for different choices of $S,\varepsilon$.}\label{fig:7x7hole_center}
\end{figure*}

We compute the probability of a hole in the site $j$ at time $\tau$
\begin{equation}
    p_\mathrm{hole, j}(\tau)= 1 - \langle n_{j,\uparrow}(\tau)\rangle - \langle n_{j,\downarrow}(\tau)\rangle +  \langle n_{j,\uparrow}(\tau)n_{j,\downarrow}(\tau)\rangle.
\end{equation}
In terms of Majorana strings, the relevant observable is
\begin{equation}
    O_\mathrm{hole, j} = \frac14 (1- \mu(v_{j_{\uparrow}j^\prime_\uparrow}) -\mu(v_{j_{\downarrow}j^\prime_\downarrow}) - \mu(v_{j_{\uparrow}j^\prime_\uparrow j_{\downarrow}j^\prime_\downarrow})),
\end{equation}
where we used the notation $j_\sigma,j^\prime_\sigma$ to indicate the indices corresponding to site $j$ and spin $\sigma=\uparrow,\downarrow$.

To first validate our 2D results, we perform simulations on a $3\times3$ lattice, for which exact diagonalization is still feasible. The results for the hole probability in the center $p_\mathrm{hole, 5}(\tau)$ and on a diagonally adjacent site $p_\mathrm{hole, 9}(\tau)$ are shown in Figures~\ref{fig:3x3holecenter} and \ref{fig:3x3holetr} in  Appendix~\ref{app:additional_experiments}, where we observe a fast convergence in the truncation parameters $S,\varepsilon$. In particular, for $S \geq 6$ and $\varepsilon \leq 10^{-5}$, we reach accurate predictions.

We now turn to a more demanding simulation on a $7\times7$ lattice (OBC) in the strongly correlated regime $U/t = 8.72$, relevant for cuprate physics. This setup challenges any classical method and provides a stringent benchmark for assessing our algorithm's accuracy against modern analog quantum simulators.
In Figure~\ref{fig:7x7hole_center} we show the hole probability $p_\mathrm{hole, center}(\tau)$ in the middle of the lattice, and compare it to the experimental results reported in Ref.~\cite{Ji2021PRX}. For $S=4,6$ and the more accurate choices of $\varepsilon$, we observe qualitative agreement between the numerical results and the experimental values, with a non-negligible dependence on $\varepsilon$ at times $\tau/t \gtrsim 1$.

\begin{figure*}
    \includegraphics[width=\textwidth]{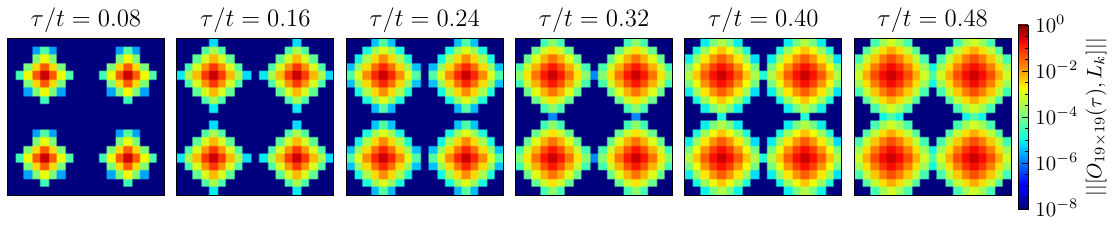}
    \caption{Expansion of $O_\mathrm{19\times19}(\tau)$ on the $19\times19$ lattice determined by the local commutators $L_k$ \eqref{eq:Lk}. The Majorana truncations employed are $S=4, \varepsilon =10^{-6}$. Even at the short timescale of $\tau/t \approx 0.5$, the observables of the 4 holes have overlapping support, and hence cannot be treated as independent and commuting observables.}\label{fig:19x19LR}
\end{figure*}

The comparison is affected by finite-size effects: in the experiment, four $7\times 7$ active regions were embedded in a larger $\sim 400$-site system to improve sampling statistics, whereas our simulation treats a single isolated region. 
As shown by the commutators in Fig.~\ref{fig:7x7LR} in Appendix~\ref{app:additional_experiments}, the information $O_{\mathrm{hole,~center}}(\tau)$ already spreads across the entire lattice for $\tau/t < 1$, indicating non-negligible overlap between neighboring regions. 
Simulations on a larger $19\times19$ lattice with four separated holes (i.e.\ similar to the experiment), defined as $O_{19\times19} = \sum_{R\in \text{regions}} O_{\text{hole,center of }R}$ confirm this intuition: by $\tau/t \approx 0.4$ the supports of these observables begin to overlap, implying that simultaneous measurements of multiple holes cannot be treated as fully independent processes, see Fig.~\ref{fig:19x19LR}.

\section{Summary and outlook}
We presented a Heisenberg-picture simulator for interacting local lattice fermion Hamiltonians based on a closed calculus of Majorana strings and a Trotter-consistent truncation that caps the growth of unpaired Majoranas. 
MP is expected to be exactly solvable in two regimes: (i) for fermionic Clifford gates and (ii) for Gaussian dynamics (provided that the initial observable can be written using strings for low-weight sectors). Moving away from such exact regimes, we expect that MP remains a very accurate classical tool for small ``dopings'', namely when one introduces in the circuit (i) non-(fermionic)-Clifford gates or (ii) non-Gaussian operations. Determining the scaling of the error as a function of such ``dopings'' is left as future work. 
Going beyond analytical guarantees, we numerically demonstrated that the method works well on a wide range of free and interacting models: from (i) Gaussian dynamics, where weight conservation keeps the evolution in a tractable subspace; to (ii) 1D Fermi-Hubbard quenches of both Fock states and variational states, where results are reliable to time scales beyond those of fermionic tensor networks; and (iii) a challenging 2D setup relevant to cuprates, where our hole-dynamics agree qualitatively with state-of-the-art ultracold-atom experiments and expose finite-size effects through Lieb-Robinson diagnostics. Furthermore, we demonstrate how the method can be combined with variational representations of the initial state.
Together, these results show that controlled truncations in the Majorana basis provide useful, systematically improvable classical estimates of real-time observables in two-dimensional fermionic systems at short to intermediate time scales.

\begin{acknowledgements}
    The authors would like to thank Manuel Rudolph for tips, tricks, and assistance related to the implementation, and for feedback on the manuscript. We thank Bart Andrews and Adrián Pérez-Salinas for insightful discussions.
    
    All tensor-network computations were carried out using ITensor~\cite{ITensor, ITensor-r0.3} and ITensorMPS. Exact dynamics references were generated with an extension of NetKet~\cite{vicentini2022netket}.
    
    \textit{During the course of this work, we became aware of concurrent work in Ref.~\cite{miller2025simulationfermioniccircuitsusing}, developed independently with different goals. Our work introduces a Heisenberg Majorana-propagation simulator for dynamics, validated on large 2D Fermi–Hubbard quenches, while Ref.~\cite{miller2025simulationfermioniccircuitsusing} applies a related method to fermionic circuits for molecular ground-state preparation. We thank its authors for helpful discussions clarifying the relation between the approaches.}
\end{acknowledgements}

\section*{Code availability}
Pauli and Majorana propagation are so closely related from a procedural point of view, that it is possible to implement Majorana propagation as a variant of the \texttt{PauliPropagation.jl} \cite{rudolph2025paulipropagationcomputationalframework} library, thereby taking advantage of most of the established routines. MP is available as open source Julia library at  \texttt{MajoranaPropagation.jl} \cite{MP.jl}.

\newpage

\clearpage
\appendix
\onecolumngrid

\section{Majorana representation}\label{app:decompositions}
We use the definition of Majorana strings \eqref{eq:gamma_def} to write the most common terms in fermionic Hamiltonians. We start by writing the creation and annihilation operators 
\begin{equation}
f_i=\frac{1}{2}\left(\gamma_i+i \gamma_i^{\prime}\right) \quad f_i^{\dagger}=\frac{1}{2}\left(\gamma_i-i \gamma_i^{\prime}\right).
\end{equation}
Number operators are then written as
\begin{equation}
n_i=f_i^\dagger f_i=\frac{1}{2}\left(1+i \gamma_i \gamma_i^{\prime}\right)\equiv\frac{1}{2}\left(1+\mu\left(v_{ii^\prime}\right)\right).
\end{equation}
The operator $i \gamma_i \gamma_i^{\prime}\equiv \mu(v_{ii^\prime})$ is a Majorana string, i.e. it satisfies \eqref{eq:ms_def}. Repulsion terms of the form $n_i n_j$ are thus 
\begin{equation}\label{eq:ms_repulsion}
n_i n_j = \frac14\left(1 + \mu(v_{ii^\prime}) +\mu(v_{jj^\prime}) - \mu(v_{ii^\prime jj^\prime})\right),
\end{equation}
where (assuming $i<j$) we defined the Majorana string $\mu(v_{ii^\prime jj^\prime}) \equiv \gamma_i\gamma_i^\prime\gamma_j\gamma_j^\prime$, since $ \mu(v_{ii^\prime})\mu(v_{jj^\prime}) = - \gamma_i\gamma_i^\prime\gamma_j\gamma_j^\prime$.
All the terms in \eqref{eq:ms_repulsion} commute with each other, hence the exponential of repulsion terms is written as
\begin{equation}
e^{-i\theta n_i n_j} = e^{-i\theta/4}e^{-i\theta/4 \mu(v_{ii^\prime})}e^{-i\theta/4 \mu(v_{jj^\prime})}e^{i\theta/4 \mu(v_{ii^\prime jj^\prime})}.
\end{equation}

Hopping terms are given by
\begin{equation}
f_i^{\dagger} f_j+f_j^{\dagger} f_i=\frac{1}{2}\left(i\gamma_i \gamma_j^{\prime}-i\gamma_i^{\prime} \gamma_j\right)\equiv \frac12\left(\mu(v_{ij^\prime})-\mu(v_{i^\prime j})\right),
\end{equation}
and the two operators $i\gamma_i \gamma_j^{\prime}\equiv \mu(ij^\prime), i\gamma_i^{\prime} \gamma_j\equiv\mu(i^\prime j)$ satisfy \eqref{eq:ms_def}. Furthermore, since $\mu(ij^\prime)$ and $\mu(i^\prime j)$ commute, we write the exponential of the hopping operator as
\begin{equation}
e^{-i \theta\left(f_i^{\dagger} f_j+f_j^{\dagger} f_i\right)}=e^{-\frac{i \theta}{2}\left(\mu\left(v_{i j^{\prime}}\right)-\mu\left(v_{i^{\prime} j}\right)\right)} = e^{-\frac{i \theta}{2} \mu\left(v_{i j^{\prime}}\right)} e^{\frac{i \theta}{2} \mu\left(v_{i^{\prime} j}\right)}
\end{equation}

\subsection{Multiplicative factors}
We give the explicit form of the prefactor $\zeta=\zeta\left(v, \tilde v\right)\in \{\pm 1, \pm i\}$ appering in the ``closeness condition'' \eqref{eq:ms_group}. We have \cite{bettaque2025structuremajoranacliffordgroup}
\begin{equation}\label{eq:pre1_app}
\begin{split}
    \zeta\left(v, \tilde v\right) &=(-1)^{v^T \omega_{\mathrm{L}} \tilde v+f\left(v, \tilde v\right)}i^{v^T \omega \tilde v} \\ &\equiv (-1)^{g(v,\tilde v)}i^{v^T \omega \tilde v}
\end{split}
\end{equation}
for 
\begin{equation}\label{eq:pre2_app}
\begin{aligned}
f\left(v, \tilde v\right) = & \left(v^T \omega_L v\right)\left(\tilde v^T \omega_L \tilde v\right) \\
& +v^T \omega \tilde v\left(v^T \omega_L v+\tilde v^T \omega_L \tilde v+1\right),
\end{aligned}
\end{equation}
where all operations in \eqref{eq:pre1_app}, \eqref{eq:pre2_app} are again to be understood as $\operatorname{mod}2$.

\subsection{Commutation relations with Majorana binary vectors}

Majorana strings are uniquely described in terms of their binary vector $v\in \{0,1\}^{2N}$, see \eqref{eq:ms_def}. It is therefore worth investigating how to write operator expressions, e.g.\ the commutation relations, in terms of operations on binary vectors. In particular, we are interested in expressions to evaluate $v^T \omega u$, determining if $v$ and $u$ commute or anticommute \eqref{eq:ms_commrel}, and $v^T \omega_L u$, which is required to compute the appropriate prefactors in \eqref{eq:ms_group}. The explicit form of the two matrices $\omega, \omega_L$ is~\cite{bettaque2025structuremajoranacliffordgroup} 
\begin{equation}
   \omega_L\equiv\left(\begin{array}{ccccc}
0 & 0 & \cdots & 0 & 0 \\
1 & 0 & \cdots & 0 & 0 \\
\vdots & \vdots & \ddots & \vdots & \vdots \\
1 & 1 & \cdots & 0 & 0 \\
1 & 1 & \cdots & 1 & 0
\end{array}\right)
\end{equation}
and 
\begin{equation}
    \omega \equiv \omega_L + \omega_L^T=\left(\begin{array}{ccccc}
0 & 1 & \cdots & 1 & 1 \\
1 & 0 & \cdots & 1 & 1 \\
\vdots & \vdots & \ddots & \vdots & \vdots \\
1 & 1 & \cdots & 0 & 1 \\
1 & 1 & \cdots & 1 & 0
\end{array}\right).
\end{equation}
We then have
\begin{equation}\label{eq:bitstrings_comm}
    v^T\omega u = \sum_k v_k (\omega u)_k = \sum_k v_k (w(u)-u_k) = w(v)w(u) - w(v \odot u),
\end{equation}
where $\odot$ is elementwise multiplication. As always, all results have to be taken $\operatorname{mod}2$. Equation \eqref{eq:bitstrings_comm} gives a quick way to decide if two strings commute or not: multiply the weights of the two strings and subtract the number of indices where they overlap. If the result is even, they commute. If it is odd, they anticommute. Consider as an example $\mu(v_{ii^\prime}) = i\gamma_i\gamma^\prime_i$ (arising from a number term) and $\mu(v_{ij^\prime}) = i\gamma_i \gamma_j^{\prime}$ (arising from a hopping term). Both have weight $2$, and they overlap on $1$ index ($\gamma_i$), hence they anticommute since $2\cdot 2 -1=3$ is odd.

\subsection{Interplay between truncation and charge conservation}
As we reported in the main text, the most straightforward approach to applying truncations in MP is to truncate after each Majorana rotation $\exp(-i\theta \mu(v)/2)$. However, we are interested in simulating the dynamics of fermionic physical Hamiltonians, which normally require multiple Majorana rotations to express basic fermionic operators. For example we know from \eqref{eq:ms_hopping} that a hopping gate can be written as 
\begin{equation}
    \exp(-i \theta (f_i^{\dagger} f_j+f_j^{\dagger} f_i)) =\exp(-i \theta \left(\mu(v_{ij^\prime})-\mu(v_{i^\prime j})\right)/2) = \exp(-i \theta \mu(v_{ij^\prime})/2)\exp(+i \theta \mu(v_{i^\prime j})/2).
\end{equation}
Last equality holds since $[\mu(v_{ij^\prime}), \mu(v_{i^\prime j})]=0$. On the other hand the cumulative density on sites $i$ and $j$, $n_i+n_j = 1+\left(\mu\left(v_{jj^\prime}\right) + \mu\left(v_{ii^\prime}\right)\right)/2$ does not commute with $\mu(v_{ij^\prime}), \mu(v_{i^\prime j})$ individually, but only with the full hopping term $\mu(v_{ij^\prime})- \mu(v_{i^\prime j})$. For MP this means that ``charge conservation''
\begin{equation}
    e^{i \theta \mu(v_{ij^\prime})/2}e^{-i \theta \mu(v_{i^\prime j})/2}(n_i+n_j) e^{+i \theta \mu(v_{i^\prime j})/2}e^{-i \theta \mu(v_{ij^\prime})/2} = n_i + n_j
\end{equation}
holds only if the truncation does not affect the application of the two gates. 

\begin{figure*}
    \includegraphics[width=\textwidth]{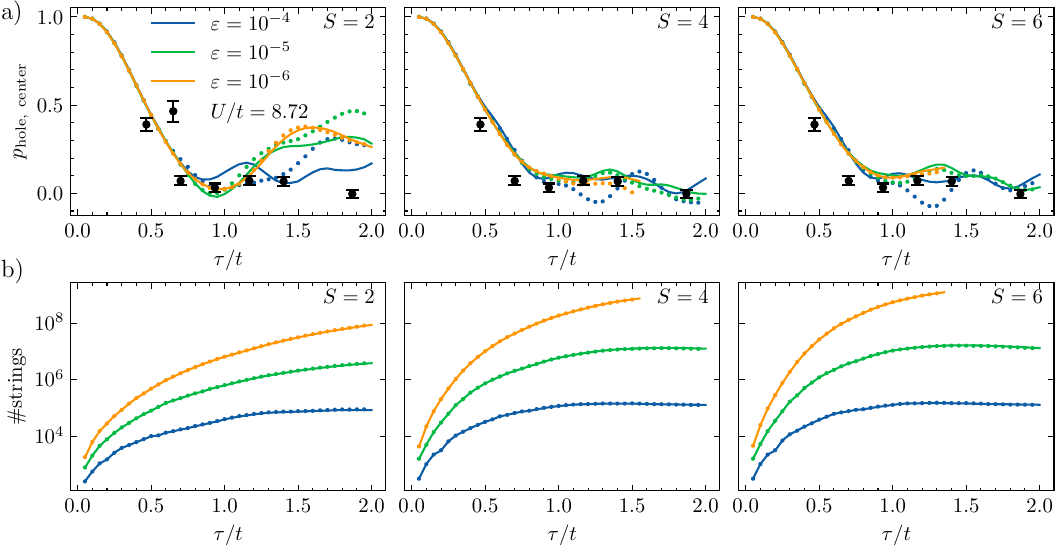}
    \caption{a) Comparison between truncating after each Majorana rotation (solid line) and truncating after each fermionic gate (dotted line). For the same setup as Figure \ref{fig:7x7hole_center}.  b) The number of strings in the observables for the two truncation schemes.}\label{fig:7x7hole_center_extra}
\end{figure*}

We hence study the effect of only truncating after applying the full fermionic rotation. The comparison for the hole dynamics is in Figure \ref{fig:7x7hole_center_extra}. We notice that this choice matters most for the less-accurate truncations, but the number of strings produced by the two schemes is nearly identical.
A full characterization of the impact of this choice is a central topic that we aim to investigate in future work.

\section{Useful expressions}

\subsection{Overlap with Fock basis states}\label{sec:Fock_overlap}

Consider the Fock basis states \[\left|n_1 \cdots n_N\right\rangle=\left(f_N^{\dagger}\right)^{n_N} \cdots\left(f_1^{\dagger}\right)^{n_1}|0\rangle\] for $|0\rangle$ the fermionic vacuum.
With the definition of the Majorana strings \eqref{eq:ms_def}, the expectation value of $\mu(v)$ wrt to $\left|n_1 \cdots n_N\right\rangle$ is computed as 
\begin{equation}\label{eq:expect_value_initial}
    \begin{split}
&\left\langle n_1 \cdots n_N\right | \mu(v)\left|n_1 \cdots n_N\right\rangle= \\&
=(i)^{v^T \omega_L v}\langle0| f_1^{n_1} \ldots f_N^{n_N} \gamma_1^{v_1} (\gamma_1^\prime)^{v_2} \cdots \gamma_N^{v_{2 N-1}} (\gamma_N^\prime)^{v_{2 N}}\left(f_N^{\dagger}\right)^{n_N} \cdots\left(f_1^{\dagger}\right)^{n_1}|0\rangle
    \end{split}
\end{equation}
We now prove that \eqref{eq:expect_value_initial} is non-vanishing only when $v$ has fully paired $\gamma$ and $\gamma^\prime$, and furthermore that
\begin{equation}\label{eq:expect_value_final}
    \left\langle n_1 \cdots n_N\right | \mu(v)\left|n_1 \cdots n_N\right\rangle = (i)^{v^T \omega_L v}\prod_{j=1}^N \left(i (-1)^{n_j}\right)^{v_{2j-1}} \delta_{v_{2j-1}, v_{2j}} 
\end{equation}
\begin{proof}
We start by showing that $\left\langle n_1 \cdots n_N\right | \mu(v)\left|n_1 \cdots n_N\right\rangle=0$ if $v$ has unpaired $\gamma$.
Assume $v_{2j-1}=1, v_{2j}=0$. Then 
\begin{equation}
    \begin{split}
\left\langle n_1 \cdots  n_N\right | & \mu(v)\left|n_1 \cdots n_N\right\rangle=(i)^{v^T \omega_L v}\langle0| f_1^{n_1}f_2^{n_2}\ldots f_N^{n_N} \gamma_1^{v_1} \cdots (\gamma_{j-1}^\prime)^{v_{2j-2}} \gamma_j \gamma_{j+1}^{v_{2j+1}} \cdots (\gamma_N^\prime)^{v_{2 N}}\left(f_N^{\dagger}\right)^{n_N} \cdots\left(f_1^{\dagger}\right)^{n_1}|0\rangle \\
& = (i)^{v^T \omega_L v}\langle0| f_1^{n_1}f_2^{n_2}\ldots f_N^{n_N} \gamma_1^{v_1} \cdots (\gamma_{j-1}^\prime)^{v_{2j-2}} (f_j^\dagger + f_j) \gamma_{j+1}^{v_{2j+1}} \cdots (\gamma_N^\prime)^{v_{2 N}}\left(f_N^{\dagger}\right)^{n_N} \cdots\left(f_1^{\dagger}\right)^{n_1}|0\rangle. \\
    \end{split}
\end{equation}
If $n_j=1$, then the expectation value vanishes since $\left(f_j^{(\dagger)}\right)^2=0$. If $n_j=0$, we use the $f_j$ (resp. $f_j^\dagger$) to annihilate $\ket0$ (resp. $\bra0$) directly.

We now move to the ``paired'' case: $v_{2j-1}=v_{2j}$. From the definition of the Majorana operators \eqref{eq:gamma_def} we have $\gamma_j\gamma_j^\prime =\left(f_j^\dagger+f_j\right)i\left(f_j^{\dagger}-f_j\right) = 2i f_jf_j^\dagger -i$.
We can simplify the expectation value by iteratively (and starting from $j=1$) bringing the Majorana operators to the left:
\begin{itemize}
\item assume $v_1=v_2=1$, then since for $i\neq j:$ $f_jf_if_i^\dagger = f_i f_i^\dagger f_j$
\begin{equation}
    \begin{split}
&\left\langle n_1 \cdots n_N\right | \mu(v)\left|n_1 \cdots n_N\right\rangle= \\&
=(i)^{v^T 0_L v}\langle0| f_1^{n_1} (2i f_1f_1^\dagger -i)f_2^{n_2}\ldots f_N^{n_N} \gamma_2^{v_3} \cdots (\gamma_N^\prime)^{v_{2 N}}\left(f_N^{\dagger}\right)^{n_N} \cdots\left(f_1^{\dagger}\right)^{n_1}|0\rangle \\
&= \left\{\begin{array}{lr}
    -i\cdot (i)^{v^T 0_L v}\langle0| f_1f_2^{n_2}\ldots f_N^{n_N} \gamma_2^{v_3} \cdots (\gamma_N^\prime)^{v_{2 N}}\left(f_N^{\dagger}\right)^{n_N} \cdots\left(f_1^{\dagger}\right)^{n_1}|0\rangle & \mathrm{if~}n_1 = 1 \\
    i\cdot (i)^{v^T 0_L v} \langle0|f_2^{n_2}\ldots f_N^{n_N} \gamma_2^{v_3} \cdots (\gamma_N^\prime)^{v_{2 N}}\left(f_N^{\dagger}\right)^{n_N} \cdots\left(f_1^{\dagger}\right)^{n_1}|0\rangle&\mathrm{if~}n_1 = 0 \\
\end{array}\right. \\
& = i (-1)^{n_1} \cdot (i)^{v^T 0_L v} \langle0| f_1^{n_1}f_2^{n_2}\ldots f_N^{n_N} \gamma_2^{v_3} \cdots (\gamma_N^\prime)^{v_{2 N}}\left(f_N^{\dagger}\right)^{n_N} \cdots\left(f_1^{\dagger}\right)^{n_1}|0\rangle
    \end{split}
\end{equation}
\item for the case $v_1=v_2=0$ 
\[\begin{split}&\langle0| f_1^{n_1}f_2^{n_2}\ldots f_N^{n_N} \gamma_1^{v_1} \cdots (\gamma_N^\prime)^{v_{2 N}}\left(f_N^{\dagger}\right)^{n_N} \cdots\left(f_1^{\dagger}\right)^{n_1}|0\rangle =\\ &= 1\cdot \langle0| f_1^{n_1}f_2^{n_2}\ldots f_N^{n_N} \gamma_2^{v_3} \cdots (\gamma_N^\prime)^{v_{2 N}}\left(f_N^{\dagger}\right)^{n_N} \cdots\left(f_1^{\dagger}\right)^{n_1}|0\rangle\end{split}.\]
\end{itemize}
Therefore, we can write
\begin{equation}
    \begin{split}
&(i)^{v^T \omega_L v}\langle0| f_1^{n_1} \ldots f_N^{n_N} \gamma_1^{v_1}  \cdots (\gamma_N^\prime)^{v_{2 N}}\left(f_N^{\dagger}\right)^{n_N} \cdots\left(f_1^{\dagger}\right)^{n_1}|0\rangle=\\
& = \left(i (-1)^{n_1}\right)^{v_1} \delta_{v_1, v_2}(i)^{v^T \omega_L v}\langle0| f_1^{n_1} \ldots f_N^{n_N} \gamma_2^{v_3}  \cdots (\gamma_N^\prime)^{v_{2 N}}\left(f_N^{\dagger}\right)^{n_N} \cdots\left(f_1^{\dagger}\right)^{n_1}|0\rangle
    \end{split}
\end{equation}
Doing this recursively for all other sites $j=2,\cdots, N$ leads to \eqref{eq:expect_value_final}.
\end{proof}

\section{Additional experiments and complementary analyses}\label{app:additional_experiments}

\subsection{1D simulation with variational initial states}
\begin{figure*}[htb]
    \includegraphics[width=\textwidth]{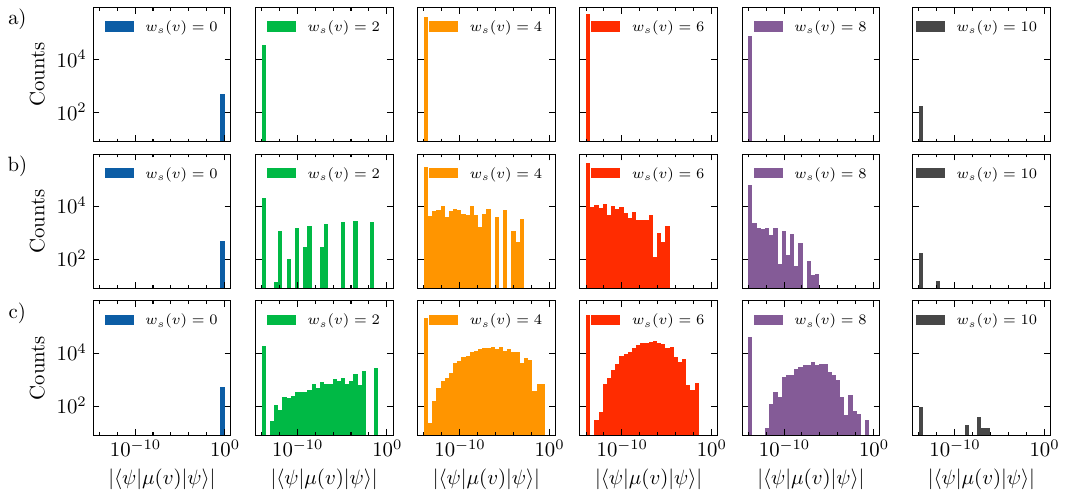}
    \caption{Distribution of the overlaps against initial states prepared with a) $U_0/t=\infty,\mu>0$, b) $U_0/t=30,\mu=50$, and c) $U_0/t=30,\mu=0.1$ as described in the main text in Section \ref{sec:1dFH}. For $U_0/t=\infty,\mu>0$ the initial state is the Fock basis state $\ket{\psi_{st}}\equiv\ket{\uparrow\downarrow\cdots\uparrow\downarrow}$, hence only strings with $w_s(v)=0$ give non-zero overlaps. Taking a finite $U_0/t$ and decreasing values of the chemical potential leads to a ground state that is a superposition of multiple Fock basis states, and hence also strings with $w_s(v)>0$ can have non-zero overlaps with such initial states.}\label{fig:mpshist}
\end{figure*}

We start by reporting the distribution of the overlaps of Majorana strings against the different ground states presented in Section \ref{sec:1dFH}. The distributions are plotted in Figure \ref{fig:mpshist}.

\subsection{Small 2D systems: benchmarks with ED}

\subsubsection{Convergence for different interaction strengths}
If the observable of interest has a low weight, evolution under Gaussian dynamics is exactly simulable with MP (see Appendix \ref{sec:gauss_dynamics}). We numerically investigate the interplay between computational hardness and non-Gaussianity.

\begin{figure*}[htb]
    \includegraphics[width=0.9\textwidth]{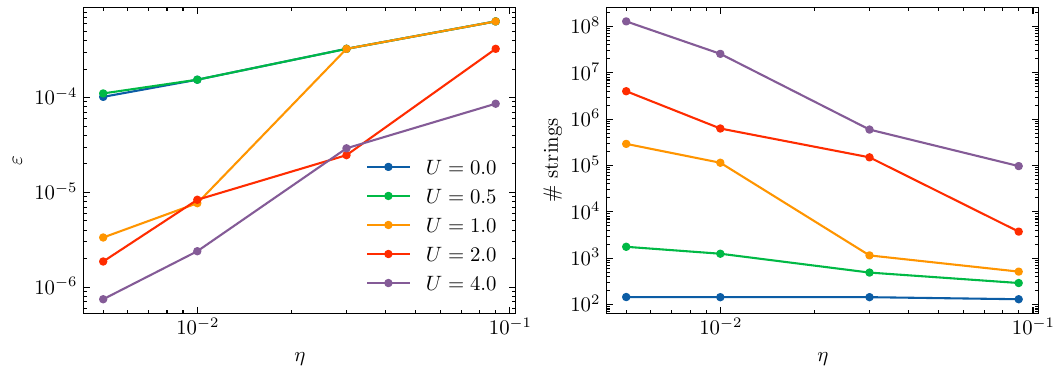}
    \caption{Left: $\varepsilon$ such that $e_\tau^\varepsilon \leq \eta$ for different $U$. Right: number of Majorana strings.}\label{fig:Uconv}
\end{figure*}

To this end, we benchmark $2$D simulations with Majorana propagation on a $4\times 3$ lattice and compare them to ED results, which are accessible at this system size. We study the dynamics of the density $n_{i,\uparrow}$ at site $i=6$ for different choices of interaction strengths $U\in\{0, 0.5, 1, 2, 4\}$. To quantify how challenging the simulation is, we fix a target accuracy $\eta$, and we simulate the system until a time $\tau$ at different coefficient truncations $\varepsilon$ until we reach 
\begin{equation}\label{eq:funky}
    \argmax_{\varepsilon} e_\tau^\varepsilon \leq \eta,
\end{equation}
where $e_\tau^\varepsilon$ is the maximal error of MP at coefficient truncation $\varepsilon$ compared to a reference solution for all times $\tau_j \leq \tau$
\begin{equation}
    e_\tau^\varepsilon := \max_{\tau_j \leq \tau} |\langle O(\tau_j)\rangle_\mathrm{MP,\varepsilon} - \langle O(\tau_j)\rangle_\mathrm{ref}|.
\end{equation}
For our numerical simulation, we select a checkerboard initial state, and a timestep $\delta\tau/t=0.05$. In Figure \ref{fig:Uconv} we report the $\varepsilon$ we found by tackling \eqref{eq:funky} as a function of the target accuracy $\eta$ for different choices of $U$. No $S$ truncation was performed.

Compared to the Gaussian regime, the $U=0.5$ case achieves very similar accuracies for almost identical $\varepsilon$, at the expense of an increase in the number of strings. For the larger interaction strengths, $U=1,2,4$, we observe that much smaller $\varepsilon$ are needed to reach the target accuracy, and consequently, the number of strings is increased a lot. Additionally, we observe the trend that the higher $U$, the smaller $\varepsilon$ needs to be to obtain similar accuracies.

\subsubsection{Hole dynamics}
\begin{figure*}[htb]
    \includegraphics[width=\textwidth]{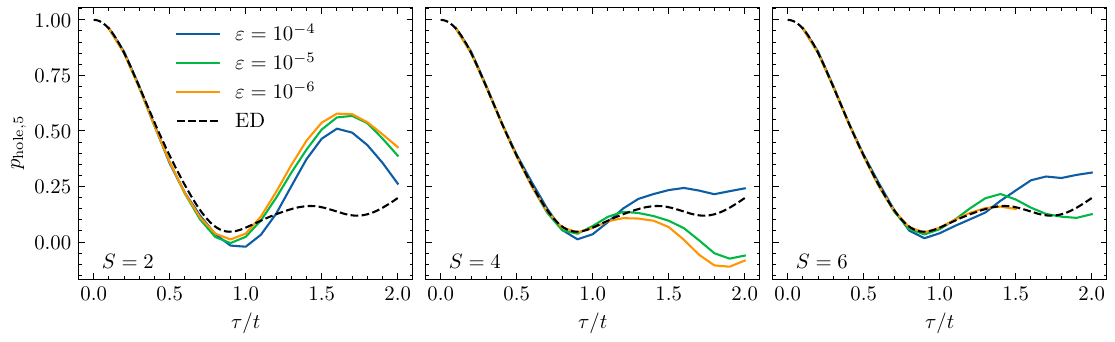}
    \caption{Dynamics of a hole in an anti-ferromagnetic background subject to $U/t=8$ on a $3\times3$ lattice. From left to right, we show various cutoffs $S=2,4,6$, and for each demonstrate the effect of the coefficient cutoff $\varepsilon$. ED results were generated with an extension of NetKet.}\label{fig:3x3holecenter}
\end{figure*}

\begin{figure*}[htb]
    \includegraphics[width=\textwidth]{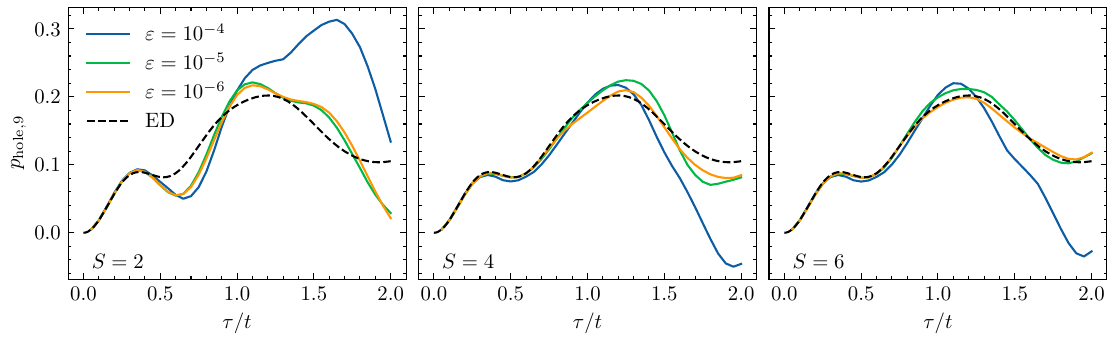}
    \caption{The hole probability of a lattice site in the corner of the lattice, for the same dynamics as Fig.~\ref{fig:3x3holecenter}. ED results were generated with an extension of NetKet.}\label{fig:3x3holetr}
\end{figure*}

We benchmark $2$D simulations with Majorana propagation on a $3\times 3$ lattice. We consider an antiferromagnetic checkerboard Fock state, with a hole at the center. Fig.~\ref{fig:3x3holecenter} and \ref{fig:3x3holetr} show the time-dependent probability of finding a hole on the central and corner sites, respectively, where we observe that for more accurate $S,\varepsilon$ the MP results approach the ED results.

\subsection{$7 \times 7$ hole-dynamics}
\subsubsection{Resource report}
\begin{figure*}[htb]
    \includegraphics[width=\textwidth]{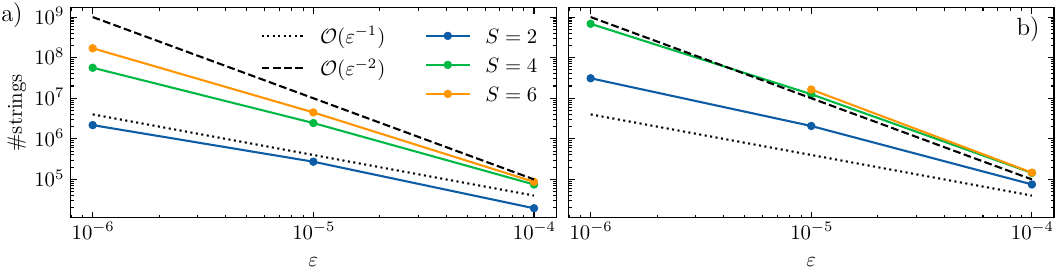}
    \caption{Number of Majorana strings as a function of $\varepsilon$ for different $S$ truncations. The data are taken from Figure \ref{fig:7x7hole_center}, at times $\tau/t=0.8$ (a) and $\tau/t=1.5$ (b).}\label{fig:7x7conv}
\end{figure*}

In this Section, we report the resources required for the most challenging simulation we performed: the hole dynamics on a $7\times 7$ spinful lattice, as shown in Figure \ref{fig:7x7hole_center}.
In Figure \ref{fig:7x7conv} we report the number of Majorana strings as a function of $\varepsilon$ at different times and $S$ truncations. In all cases, we observe power-law scaling of the number of strings, ranging $\sim\order{\varepsilon^{-1}}$ for the looser truncation the earlier times $S=2, \tau/t =0.8$ to $\sim\order{\varepsilon^{-2}}$ for the most accurate truncations at the later times $S=6, \tau/t=1.5$, matching the expected scalings of \eqref{eq:Mbound}.

In Table \ref{tab:resources} we report the runtimes and max memory requirements for each simulation. Runtimes range from less than 2 minutes for the most inaccurate calculations to around 2 weeks for the most challenging ones.

\begin{table}[]
    \centering
    \begin{tabular}{cccccc}
    \toprule
    $S$ & $\varepsilon$ & Runtime [hh:mm:ss] & Max memory [GB] &Note\\
    \midrule 
    $2$ & $10^{-4}$ & 00:01:40 & $0.61$ \\
    $2$ & $10^{-5}$ & 01:19:54 & $3.01$ \\
    $2$ & $10^{-6}$ & 41:19:16& $68.85$ \\
    \midrule
    $4$ & $10^{-4}$ & 00:03:36 & $0.98$ \\
    $4$ & $10^{-5}$ & 05:06:15 & $8.31$ \\
    $4$ & $10^{-6}$ & 217:09:40&  $466.27$ & Stopped at $\tau/t=1.55$\\
    \midrule
    $6$ & $10^{-4}$ & 00:04:12 & $0.52$ \\
    $6$ & $10^{-5}$ & 07:50:58 & $9.72$ \\
    $6$ & $10^{-6}$ & 349:21:59&  $589.45$ &Stopped at $\tau/t=1.35$\\
 \bottomrule
    \end{tabular}
    \caption{Resources for simulating the $7\times 7$ hole dynamics reported in Figure \ref{fig:7x7hole_center}.}
    \label{tab:resources}
\end{table}

\subsubsection{Convergence in $S$}
\begin{figure*}
    \includegraphics[width=0.6\textwidth]{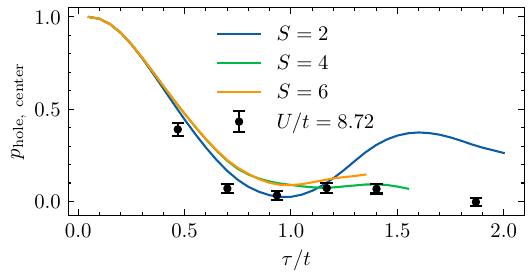}
    \caption{Results of the $7\times7$ hole-dynamics for $\varepsilon=1\cdot 10^{-6}$ and various $S$.}\label{fig:7x7S}
\end{figure*}

In Figure \ref{fig:7x7S} we report the results presented in Figure \ref{fig:7x7hole_center} for $\varepsilon=1\cdot 10^{-6}$ and $S=2,4,6$. Because of the strongly interacting nature of the setup, it is not possible to investigate the convergence of the numerical method beyond $\tau/t \approx 1.2$.

\subsubsection{LR plots}
\begin{figure*}[htb]
    \includegraphics[width=\textwidth]{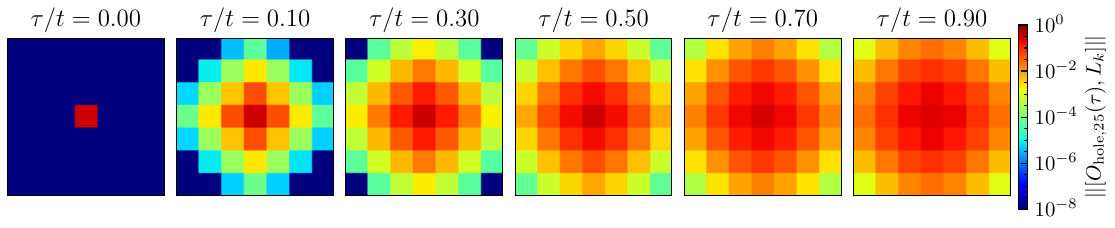}
    \caption{Expansion of $O_\mathrm{hole, j}(\tau)$ on the $7\times7$ lattice determined by the local commutators $L_k$ \eqref{eq:Lk}. The Majorana truncations employed are $S=4, \varepsilon =10^{-6}$. We observe that already at $\tau/t < 1$ the observable has spread to all $49$ sites of the lattice.}\label{fig:7x7LR}
\end{figure*}

In Figure \ref{fig:7x7LR} we display the LR lightcone for the hole dynamics experiment of Figure \ref{fig:7x7hole_center}.

\section{Gaussian fermion dynamics}\label{sec:gauss_dynamics}
Majorana propagation can efficiently simulate Gaussian fermion dynamics of observables with a low (or very high) Majorana weight. We demonstrated this in the main text, starting from the viewpoint of the branching structure for Majorana propagation in Eq.~\eqref{eq:mp_splitting} for the unitaries in the Trotterized dynamics. Here, we will demonstrate that Majorana strings indeed maintain their weight during the evolution in a more general way, to complement the discussion in the main text.

First, we introduce some notation. Consider, for simplicity of notation, a quadratic Hamiltonian that is also charge-conserving. The results can be easily generalized. We have
\begin{align}
    H &= h_{ij} f_i^\dagger f_j \\
    &= e_i d_i^\dagger d_i
\end{align}
where 
\begin{align}
    h_{ij} &= [V]_{ik} e_k [V^\dagger]_{kj} = [V E V^\dagger]_{ij}
    \label{eq:gaussian_diagonalization} \\
    d_i &= [V^\dagger]_{ik} f_k  \\
    f_i &= V_{ij} d_j
\end{align}
where we denote $[V^\dagger]_{ik} = V^*_{ki}$ and $E = \diag e_i$.
The dynamics are diagonal in the $d$ operators
\begin{align}
    d_i(t) &= e^{-i e_i t} d_i
\end{align}
and hence, with the single-particle propagator
\begin{align}
U(t) = e^{-iht} = V e^{-iEt} V^\dagger
\end{align}
we can write the dynamics of the ladder operators
\begin{align}
    f_i(t) &= U_{ij}(t) f_j \\
    f_i^\dagger(t) &= [U^\dagger(t)]_{ji} f_j^\dagger  \\
    &= U_{ij}^*(t) f_j^\dagger
\end{align}
For the Majorana modes, we have
\begin{align}
    \gamma_j(t) &= f_j^\dagger(t)+f_j(t) \\
    &= U_{jk}^*(t) f_k^\dagger + U_{jk}(t) f_k \\
    &= \frac12 \left[U_{jk}^*(t) (\gamma_k - i \gamma_k') + U_{jk}(t) (\gamma_k + i \gamma_k') \right] \\
    &= \frac12 \left[(U_{jk}(t) + U_{jk}^*(t))\gamma_k +i (U_{jk}(t) - U_{jk}^*(t)) \gamma_k' \right] 
\end{align}
and similarly
\begin{align}
\gamma_j'(t) &= i\left(f_j^\dagger(t)-f_j(t)\right) \\
&= \tfrac12 \left[ -i(U_{jk}(t) - U_{jk}^*(t)) \gamma_k
+ (U_{jk}(t)+U_{jk}^*(t)) \gamma_k' \right].
\end{align}

In short
\begin{align}
\gamma_j(t)   &= R_{jk} \gamma_k - I_{jk} \gamma_k',\\
\gamma_j'(t)  &= I_{jk} \gamma_k + R_{jk}\gamma_k'.
\end{align}
where we introduced
\begin{align}
R_{ij} &= \mathrm{Re}[U(t)]_{ij}, \qquad
I_{ij} = \mathrm{Im}[U(t)]_{ij}.
\end{align}
such that the unitarity of $U$ yields the constraints
\begin{align}
[U(t)]_{ik}[U^\dagger(t)]_{kj} &= \delta_{ij}
\quad\Rightarrow\quad
\begin{cases}
R_{ik} R_{jk} + I_{ik} I_{jk} = \delta_{ij},\\[6pt]
R_{ik} I_{jk} - I_{ik} R_{jk} = 0.
\end{cases}
\end{align}
For a time-dependent Majorana product, we then get for $i \neq j$
\begin{align}
\gamma_i(t)\gamma_j(t)
&=\sum_k\left[ R_{ik}R_{jk} + I_{ik}I_{jk} \right] - \left[R_{ik} I_{jk} - I_{ik} R_{jk}\right] \gamma_k\gamma_k^\prime \nonumber \\
 &\quad + \sum_{k\neq \ell} \Big[
   R_{ik}R_{j\ell}\gamma_k\gamma_\ell
 - R_{ik}I_{j\ell}\gamma_k\gamma_\ell'
 - I_{ik}R_{j\ell}\gamma_k'\gamma_\ell
 + I_{ik}I_{j\ell}\gamma_k'\gamma_\ell'
 \Big] \\
&= \sum_{k\neq \ell} \Big[
   R_{ik}R_{j\ell}\gamma_k\gamma_\ell
 - R_{ik}I_{j\ell}\gamma_k\gamma_\ell'
 - I_{ik}R_{j\ell}\gamma_k'\gamma_\ell
 + I_{ik}I_{j\ell}\gamma_k'\gamma_\ell'
 \Big].
\end{align}
where we used unitarity to remove the diagonal contribution given by the first line. The latter would form a problematic term that might otherwise generate Majorana strings of different weights.

We similarly get
\begin{align}
\gamma_i(t)\gamma_j'(t)
&=\sum_k \big(R_{ik}R_{jk}+I_{ik}I_{jk}\big)\gamma_k\gamma_k' \\
&\quad + \sum_{k\neq \ell} \Big[
   R_{ik}I_{j\ell}\gamma_k\gamma_\ell
 + R_{ik}R_{j\ell}\gamma_k\gamma_\ell'
 - I_{ik}I_{j\ell}\gamma_k'\gamma_\ell
 - I_{ik}R_{j\ell}\gamma_k'\gamma_\ell'
 \Big].
\end{align}
Hence, we find that $\gamma_i(t) \gamma_j(t)$ and $\gamma_i(t) \gamma_j'(t)$ remain in the subspace generated by the set
\begin{align}
    \left\{ \gamma_k \gamma_\ell, \gamma_k' \gamma_\ell' \right\}_{k\neq \ell} \cup \{\gamma_k \gamma_\ell', \gamma_k' \gamma_\ell\}_{k,l}
\end{align}
Hence, we showed explicitly that any weight $2$ Majorana string (see Eq.~\eqref{eq:ms_weight}) stays in a subspace spanned by weight $2$ Majorana strings when subjected to the Gaussian dynamics of quadratic Hamiltonians. This result can trivially be generalized to any even-parity (or even weight) Majorana string, by combining the above results on all sets of two Majoranas. Therefore, we can conclude that \emph{any weight $w$ Majorana string remains in a subspace spanned by weight $w$ Majorana strings when subjected to the Gaussian dynamics of quadratic Hamiltonians.
}
This implies that, as long as the subspace of a given weight is small enough, MP can simulate the dynamics of a given observable efficiently. For the spatially local observables considered here, this generally holds. In the more general case, the dimension of the subspace of weight $w$ is $D = \binom{2N}{w}$ where $N$ is the number of fermionic modes.

\FloatBarrier

\section{Gaussian initial states}\label{app:gaussian_initial}
We consider the expectation value of an operator $O$ under a time evolution $U(\tau)$, starting from an initial state
\begin{align}
    \ket{\psi} = W \ket{n_1 \cdots n_N},
\end{align}
where $\ket{n_1 \cdots n_N}$ again denotes a simple Fock state, see Eq. \eqref{eq:fock_state}. If the columns of $W$ contain the one-body eigenstates ordered by increasing single-particle energies, this state represents the Hartree–Fock (HF) ground state.

When $\ket{\psi}$ is the ground state of a quadratic Hamiltonian, it can be advantageous to absorb the transformation $W$ into the time evolution of the operator—rather than explicitly representing $\ket{\psi}$ as a potentially high bond-dimension tensor network. In this way, the expectation value can be rewritten as
\begin{align}
    \mel{n_1 \cdots n_N}{W^\dagger U^\dagger(\tau) O U(\tau) W}{n_1 \cdots n_N}.
\end{align}
The unitary $W$ associated with the quadratic Hamiltonian is obtained through a one-body diagonalization (see Section~\ref{sec:gauss_dynamics}, Eq.~\eqref{eq:gaussian_diagonalization}), with the columns of the diagonalizing matrix ordered according to energy. Specifically, if
\begin{align}
    d_i = [V^\dagger]_{ik} f_k,
\end{align}
defines the single-particle transformation from the original fermionic operators $f_k$ to the diagonal modes $d_i$, then there exists an anti-Hermitian matrix $K$ such that $e^K = V$. The corresponding many-body rotation is the second-quantized unitary
\begin{align}
    W = \exp\left(\sum_{ij} K_{ij} f_i^\dagger f_j\right),
\end{align}
which implements the transformation $W f_i W^\dagger = \sum_j V_{ij} f_j$ and thus generates the desired Gaussian state.

\section{Excitation error}\label{app:excitation-error}
Here we report a brief summary of the excitation error introduced in \cite{dai2025fermionic}. For each site $i$, we define $n_i$ as the occupation number \textit{relative to the ground state}. We introduce $n_{\mathrm{tot}}=\sum_i n_i$, and we select excitation sites according to their contribution to the total occupation number $n_{\mathrm{tot}}$
\begin{equation}\label{eq:exc_sites}
    \mathcal I_e=\left\{i \mid \sum_{n_j<n_i} n_j<(1-\eta) n_{\mathrm{tot}}\right\}.
\end{equation}
The excitation error is then
\begin{equation}
    \text { Excitation error }=\sum_{i \in \mathcal I_e} \Delta n_i / n_{\text {tot }},
\end{equation}
where $\Delta n_i = \abs{n_{i,\text{sim}}-n_{i,\text{exact}}}$.

\bibliography{biblio}

\end{document}